\documentclass[aps,prl,twocolumn,superscriptaddress]{revtex4}

\usepackage{mathrsfs}
\usepackage{graphicx}
\usepackage{bm}
\usepackage{threeparttable}
\usepackage{amsmath,amssymb}
\usepackage{color}
\usepackage{epstopdf}
\usepackage[hidelinks]{hyperref}
\usepackage{changes}

\begin{document}
%\preprint{}



\title[%
cell slow model]{Interfacial Stability in Tensionless Phase-Separated Quorum-Sensing Systems %\\This line break forced with
}%

\author{Zihao Sun}
\thanks{These authors contributed equally to this work.}
\affiliation{Beijing National Laboratory for Condensed Matter Physics and Laboratory of Soft Matter Physics, Institute of Physics, Chinese Academy of Sciences, Beijing 100190, China}
\affiliation{School of Physical Sciences, University of Chinese Academy of Sciences, Beijing 100049, China}
\author{Longfei Li}
\thanks{These authors contributed equally to this work.}
\affiliation{Beijing National Laboratory for Condensed Matter Physics and Laboratory of Soft Matter Physics, Institute of Physics, Chinese Academy of Sciences, Beijing 100190, China}
\author{Fangfu Ye}
\email[fye@iphy.ac.cn]{}
\affiliation{Beijing National Laboratory for Condensed Matter Physics and Laboratory of Soft Matter Physics, Institute of Physics, Chinese Academy of Sciences, Beijing 100190, China}
\affiliation{School of Physical Sciences, University of Chinese Academy of Sciences, Beijing 100049, China}
\affiliation{Wenzhou Institute, University of Chinese Academy of Sciences, Wenzhou, Zhejiang 325001, China}
\affiliation{Oujiang Laboratory (Zhejiang Lab for Regenerative Medicine, Vision and Brain Health), Wenzhou, Zhejiang 325000, China}
%\affiliation{Songshan Lake Materials Laboratory, Dongguan, Guangdong 523808, China}
\author{Mingcheng Yang}
\email[mcyang@iphy.ac.cn]{}
\affiliation{Beijing National Laboratory for Condensed Matter Physics and Laboratory of Soft Matter Physics,
Institute of Physics, Chinese Academy of Sciences, Beijing 100190, China}%
\affiliation{School of Physical Sciences, University of Chinese Academy of Sciences, Beijing 100049, China}

%\thanks{}


\date{\today}

%\pacs{}% insert suggested PACS numbers in braces on next line
\begin{abstract}
Interfacial phenomena of motility-induced phase separation of active particles challenge our conventional understanding of phase coexistence. Despite the ubiquity of nonmechanical communication couplings among real active particles, most works on active interface have concentrated on active Brownian systems with steric interparticle interactions. Here, we study the interfacial behavior of phase-separated active particles interacting solely via quorum-sensing communications using both theory and simulations. Strikingly, we find that the quorum-sensing active system exhibits vanishing mechanical surface tension but nonzero effective capillary surface tension. We further demonstrate that the mechanical equilibrium of the tensionless interface is sustained by polarization force at the interface; while its dynamics is governed by the surface stiffness, which arises from tangential particle flux induced by local interfacial deformation. Our work reveals the fundamental distinction between mechanical and capillary surface tensions in active matter and paves the way for future exploration of active interface phenomena.
\end{abstract}
\maketitle %\maketitle must follow title, authors, abstract and \pacs

\emph{Introduction}\textemdash Motility-induced phase separation (MIPS) refers to spontaneous separation of active particles into dense and dilute phases in the absence of attractive interactions~\cite{tailleur2008statistical,cates2015motility}. MIPS is one of the most prominent phenomena in active matter, and has been extensively investigated in diverse active systems~\cite{theurkauff2012dynamic,fily2012athermal,redner2013structure,buttinoni2013dynamical,
stenhammar2013continuum,bialke2013microscopic,thompson2011lattice,digregorio2018full,
omar2023mechanical}. Despite the intrinsically nonequilibrium complexity of MIPS, its underlying microscopic mechanism is straightforward: a positive feedback loop between the slowing down of active particles at high density due to steric hindrance or other factors and the slowing-induced particle accumulation~\cite{cates2015motility,solon2015active}. However, surface tension\textemdash a pivotal physical quantity in MIPS\textemdash and related interfacial phenomena, still remain elusive and controversial~\cite{bialke2015negative,paliwal2017non,solon2018generalized,
del2019interface,hermann2019non,omar2020microscopic,lauersdorf2021phase,fausti2021capillary,
li2023surface,langford2025phase}.

In their seminal work~\cite{bialke2015negative}, Bialké et al. first calculated the surface tension of phase-separated repulsive active Brownian particles (ABPs) based on the widely-accepted active pressure concept, reporting a significantly negative value. This counterintuitive finding has since stimulated considerable debate, making the quest for proper understanding of surface tension and interfacial stability in MIPS an ongoing research focus. More surprisingly, for the MIPS of the minimal ABPs, recent studies suggest that the mechanical surface tension that determines interfacial mechanics may differ from the capillary surface tension that governs interfacial fluctuations~\cite{lee2017interface,
chacon2022intrinsic,langford2024theory,langford2024mechanics}. This is in stark contrast to the case of equilibrium phase separation, where various interfacial behaviors are governed by a common surface tension~\cite{evans1979nature,rowlinson2013molecular}. Additionally, the studies on mechanical surface tension have been limited to specific ABPs coupling via direct steric interactions. This leaves broader classes of active matter\textemdash featuring ubiquitous nonmechanical communication (e.g., biochemical or visual signals)\textemdash unexplored. The main reason for this situation is that, for generic active systems, the extensively-used active pressure is not a state function and even its local definition is ambiguous~\cite{solon2015pressure,Speck2016pre,li2023surface}. Given the complexity and fundamental importance of active surface tensions, it is critical to clarify the difference between mechanical and capillary surface tensions in MIPS and to explore the interfacial properties of nonmechanically communicating active systems. 

In this Letter, we employ the recently proposed approach based on intrinsic pressure~\cite{li2023surface} – applicable to general active matter – to investigate mechanical surface tension in the MIPS of ideal quorum-sensing (QS) active systems (without steric interactions), as well as its interfacial dynamics. Here, active individuals interact solely through the QS communication, adjusting their motility in response to local density and enabling spontaneous phase separation~\cite{solon2018generalized,
bauerle2018self,fischer2020quorum,worlitzer2021motility,gnan2022critical}. We choose the ideal QS system for two main reasons. First, QS is a key nonmechanical communication strategy, crucial in physiological processes~\cite{zhu2002quorum,
parsek2005sociomicrobiology,lupp2005vibrio,diggle2007quorum} and collective motions~\cite{cates2010arrested,rein2016collective,shaebani2020computational,
gompper20202020,worlitzer2021motility,duan2023dynamical,dinelli2023non}. Second, the lack of steric interactions in the ideal QS system minimizes mechanical surface tension (originating from anisotropic stress), facilitating distinction between different types of surface tensions. Our results reveal that the QS active system exhibits a vanishing mechanical surface tension, further confirmed by analyzing the mechanical equilibrium of the droplet's surface. Moreover, we find that the tensionless interface exhibits significant surface stiffness, hence a nonzero effective capillary surface tension. This surface stiffness arises from tangential particle currents induced by local interface deformation and governs interfacial dynamics and stability, which is verified from capillary wave analysis.

\begin{figure}[tbh]
\centering 
\includegraphics[width = 0.47\textwidth]{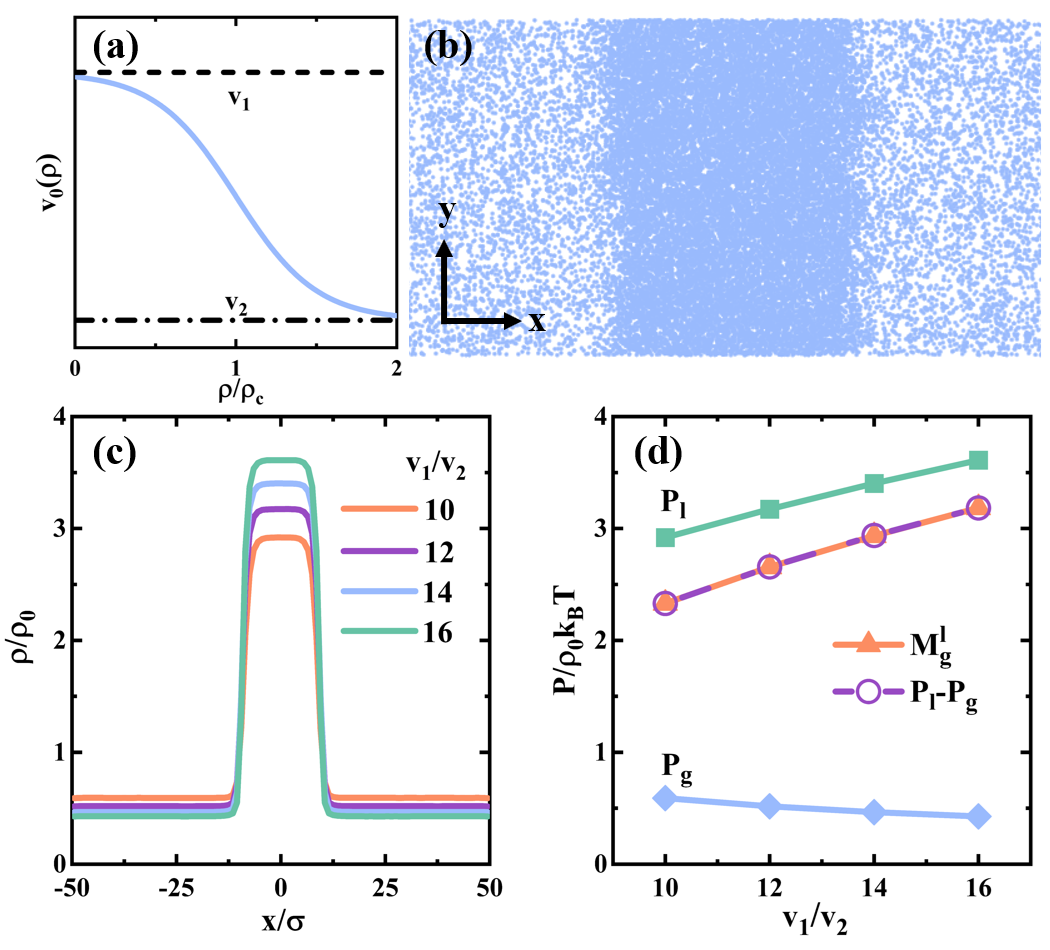}
\caption{(a) Dependence of the self-propelling velocity $v_0$ on the local particle number density, as described by Eq.~(\ref{eq2}). (b) Snapshot of the ideal phase-separated QS system with a planar interface ($L_x/L_y  = 4$, $v_1/v_2 = 10$ and $N = 40000$). (c) Steady-state density profiles $\rho(x)$, and (d) verification of the force balance condition Eq.~(\ref{eq3}) by comparing the pressure differences and polarization contributions across varying $v_1/v_2$.}
\label{fig1}
\end{figure}

\emph{Ideal QS active particle model}.\textemdash We consider a minimal active system that exhibits the MIPS due to the QS mechanisms. This system comprises $N$ nonmechanically communicating self-propelled particles in a two-dimensional box with periodic boundary conditions. The dynamics of a particle with position $\mathbf{r}_i$ and orientation $\mathbf{e}_i = (\cos \theta_i,\sin \theta_i)$ is described by the overdamped Langevin equations
\begin{align}\label{eq1}
\dot{\mathbf{r}}_i &= v_0(\rho(\mathbf{r}_i))\mathbf{e}_i + \sqrt{2D_\text{t}}\bm{\eta}_i\\
\dot{\mathbf{e}}_i &= \sqrt{2D_\text{r}}\bm{\xi}_i  \times \mathbf{e}_i,\notag
\end{align}
with $D_\text{t}$ and $D_\text{r}$ separately denoting the translational and rotational diffusion coefficients, and $\bm{\eta}_i$ and $\bm{\xi}_i$ being the Gaussian white noises of zero mean and unit variance. Here, the self-propelled velocity, $v_0(\rho(\mathbf{r}_i))$, depends on local particle number density $\rho(\mathbf{r}) = \sum_i \delta(\mathbf{r} - \mathbf{r}_i)$, which mimics the QS communication. Following the work by Solon et al.~\cite{solon2018generalized}, the velocity is taken to be 
\begin{equation}\label{eq2}
v_0(\rho) = v_1 + \frac{v_2 - v_1}{2}\left[ 1 + \tanh(2 \frac{\rho}{\rho_\text{c}} - 2) \right],
\end{equation}
with $\rho_\text{c}$ being the characteristic density. In this QS way, the particle smoothly adjusts its self-propelled velocity from a high value $v_1$ in the low-density region to a low value $v_2$ in the high-density region, as sketched in Fig.~\ref{fig1}(a). On the other hand, it is known that the active particles tend to accumulate in low-activity region~\cite{cates2015motility,solon2015active}. Consequently, a positive feedback loop is established, and ultimately the ideal QS system experiences phase separation, under the linear instability condition $dv_0/d\rho < -v_0/\rho$.

Unless otherwise specified, throughout the subsequent analyses, the system parameters are taken as $v_2 = 1$, $\rho_\text{c} = 20$, $D_\text{t} = 1$, $D_\text{r} = 1$, and the mean particle number density $\rho_0 = 16$. The implementation of the QS communication requires real-time measurements of the local density in the vicinity of the selected particle. To enhance the computational efficiency, we employ a coarse-grained particle sorting approach to quantify the local density within a squared area of dimension $\sigma = 1$, which has been widely used in mesoscale fluid simulations~\cite{malevanets1999mesoscopic,
kapral2008multiparticle} (see Supplemental Material (SM), Sec.~3.1~\cite{SupplRef} for more simulation details). 

\emph{Mechanical surface tension}\textemdash When $v_1\gg v_2$, the QS active particles spontaneously separate into a dense liquid phase of density $\rho_\text{l}$ and a dilute gas phase of density $\rho_\text{g}$ (Fig.~\ref{fig1}(b) and (c)). Note that, in this work we adopt the framework of intrinsic pressure that only comprises conventional kinetic and interacting contributions~\cite{Speck2016pre,omar2020microscopic,omar2023mechanical,
li2023surface}, as defined in passive systems. The bulk pressures of the gas and liquid phases are denoted as $P_\text{g}$ and $P_\text{l}$, respectively. In this framework, the self-propelling force, $\gamma_\text{t} v_0(\rho) \mathbf{e}_i$, is treated as an external force, so the local force balance reads,
\begin{equation}\label{eq3}
\mathbf{\nabla}\cdot\mathbf{P} = \gamma_\text{t}v_0(\rho)\mathbf{m} - \gamma_\text{t} \mathbf{J}^\rho,
\end{equation}
with $\mathbf{P}$ the pressure tensor, the polar order $\mathbf{m}(\mathbf{r}) = \sum_i \mathbf{e}_i \delta(\mathbf{r} - \mathbf{r}_i)$ and the particle flux $\mathbf{J}^\rho(\mathbf{r}) =  \sum_i \dot{\mathbf{r}}_i \delta(\mathbf{r} - \mathbf{r}_i)$. Thus, in the flux-free steady state, the mechanical equilibrium of the planar interfacial region becomes, $P_\text{l} - P_\text{g} = \int_\text{g}^\text{l}\gamma_\text{t}v_0(\rho)m_x dx = M_\text{g}^\text{l}$, where the upper and lower limits of the integration separately reach into the bulk liquid and gas. Simulation results shown in Fig.~\ref{fig1}(d) confirm this mechanical equilibrium condition of the planar interface. It should be pointed out that in the ideal QS active system without steric interactions, the local pressure is isotropic and follows the ideal-gas equation of state, $P(\mathbf{r}) = \rho(\mathbf{r}) k_\text{B}T$.

\begin{figure}[t]
\centering
\includegraphics[width = 0.47\textwidth]{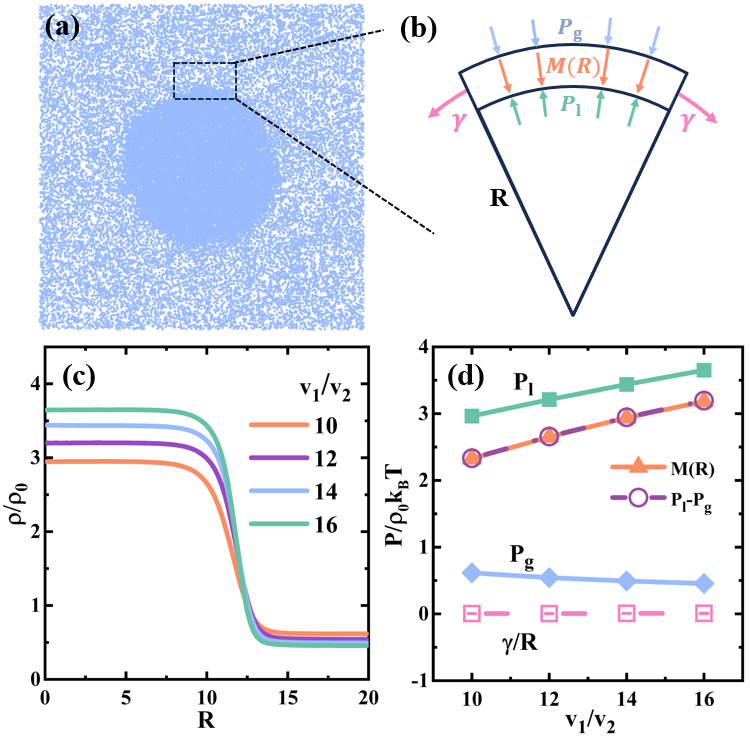}
\caption{(a) Snapshot of a steady-state droplet in a squared box with $N=40000$ and $v_1/v_2 = 10$. (b) Schematic of the force balance at the droplet surface (radius $R$). Color arrows represent: the gas pressure $P_\text{g}$ (blue), liquid pressure $P_\text{l}$ (green), polarization force contribution $M(R)$ (orange), and surface tension $\gamma$ (pink). (c) Radial density profiles $\rho(r)$ and (d) bulk pressures, pressure difference, polarization force contribution and surface tension for the droplet system with varying $v_1/v_2$.}
\label{fig2}
\end{figure}

Using the intrinsic pressure framework, the mechanical surface tension is expressed as~\cite{li2023surface}:
\begin{equation}\label{eq4}
\gamma = \int_\text{g}^\text{l} \left[ P_\text{g} \frac{\rho_\text{l} - \rho(x)}{\rho_\text{l} - \rho_\text{g}} + P_\text{l}\frac{\rho(x) - \rho_\text{g}}{\rho_\text{l}- \rho_\text{g}} - P^\text{T}(x) \right] dx.
\end{equation}
Here, the tangential pressure $P^\text{T}$ corresponds to the pressure in the $y$-direction. Strikingly, substituting this ideal-gas pressure into Eq.~(\ref{eq4}) results in a vanishing mechanical surface tension, $\gamma = 0$. This result inspires us to examine the mechanical equilibrium of a macroscopic droplet's surface (Fig.~\ref{fig2}(a)), which seems unable to be maintained by the zero mechanical surface tension. However, as demonstrated below, the interfacial polarization force is crucial for maintaining the mechanical equilibrium of the droplet.

For a droplet formed due to the MIPS, the force balance of its surface (as sketched in Fig.~\ref{fig2}(b)) obeys a generalized Young-Laplace equation~\cite{li2023surface},
\begin{equation}\label{eq5}
P_\text{l} - P_\text{g} - M(R) = \frac{\gamma}{R}.
\end{equation}
Here, $M(R)$ denotes the polarization force per unit arc length acting on the droplet’s surface, with the curvature radius $R$. To be more precise, the droplet radius $R$ is taken as the position of the equimolar dividing surface~\cite{green1960molecular,rowlinson2013molecular}, which is determined in terms of the radial density profile (Fig.~\ref{fig2}(c)). Figure~\ref{fig2}(d) demonstrates that the pressure difference $P_\text{l} - P_\text{g}$ between the liquid and gas phases is balanced by the polarization force contribution $M(R)$. Thus, Eq.~(\ref{eq5}) yields a negligible $\gamma$, aligning well with the above prediction. We point out that the phase-separated QS systems have been studied through generalized thermodynamics theory~\cite{solon2018generalized,
tjhung2018cluster}, but the resulting effective surface tensions have no mechanical interpretation. 

Notably, the MIPS interface experiences significant fluctuations, as displayed in Fig.~\ref{fig1}(b) and Video S1 (SM)~\cite{SupplRef}. However, the vanishing $\gamma$ alone is insufficient to suppress the interfacial fluctuations and to stabilize the interface. Therefore, there must be another mechanism responsible for stabilizing the MIPS interface. To explore this mechanism, we analyze the dynamics of the interface based on the fluctuating hydrodynamics of the QS active fluid, which we introduce below.

\emph{Fluctuating hydrodynamics of QS active system}\textemdash To simplify our analysis and isolate the role of activity, we set the translational noise of the active particles to zero, i.e. $D_\text{t} = 0$. Starting from Eq.~(\ref{eq1}), we derive the following fluctuating hydrodynamic equation~\cite{dean1996renormalization,cugliandolo2015stochastic,langford2024theory} (see SM, Sec.~1~\cite{SupplRef}, for detailed derivations),
\begin{align}
\frac{\partial \rho}{\partial t} &= -\bm{\nabla} \cdot \mathbf{J}^\rho,\label{eq6}\\
\frac{\partial \mathbf{m}}{\partial t} &= -\bm{\nabla} \cdot \mathbf{J}^m - D_\text{r} \mathbf{m} + \bm{\eta}^m.\label{eq7}
\end{align} 
Here, $\mathbf{J}^\rho(\mathbf{r}) = v_0(\rho)\mathbf{m}$ and $\mathbf{J}^m(\mathbf{r}) = v_0(\rho)(\rho\mathbf{I}/2 + \mathbf{Q})$ are the flux of particle and polar order, respectively, with the traceless nematic order tensor $\mathbf{Q}(\mathbf{r}) = \sum_i (\mathbf{e}_i \mathbf{e}_i - \mathbf{I}/2)\delta(\mathbf{r} - \mathbf{r}_i)$. Besides, a stochastic term $\bm{\eta}^m(\mathbf{r},t) = \sum_i \sqrt{2D_\text{r}}\bm{\xi}_i \times \mathbf{e}_i \delta(\mathbf{r} - \mathbf{r}_i)$ is incorporated to account for the polar order fluctuation. It is characterized by a zero mean $\langle \bm{\eta}^m (\mathbf{r},t) \rangle = 0$ and a variance $\langle \bm{\eta}^m (\mathbf{r},t) \bm{\eta}^m (\mathbf{r}',t')\rangle = D_\text{r}(\rho\mathbf{I} - 2\mathbf{Q})\delta(\mathbf{r} - \mathbf{r'})\delta(t - t')$.

We truncate the hydrodynamic hierarchy described in Eq.~(\ref{eq7}) by neglecting the nematic order parameter $\mathbf{Q}$, a simplification justified numerically (shown in SM, Sec.~1 and Sec.~2.2~\cite{SupplRef}). Given that the relaxation timescale of polar order ($\tau_\text{r} = 1/D_\text{r}$) is much shorter than the timescale for density field fluctuations (interfacial capillary waves), we ignore the temporal evolution of polar order in the hydrodynamic analysis. Consequently, the particle flux becomes,
\begin{equation}\label{eq8}
\mathbf{J}^{\rho} = v_0(\rho)\mathbf{m} = \frac{v_0(\rho)}{D_\text{r}}\left[-\bm{\nabla}\cdot \left(\frac{v_0(\rho)\rho}{2}\mathbf{I}\right) + \bm{\eta}^m\right].
\end{equation}
With the evolution equations solely for the density field in place, we now examine the fluctuation behavior of the MIPS interface.

\begin{figure}[htb]
\centering 
\includegraphics[width = 0.47\textwidth]{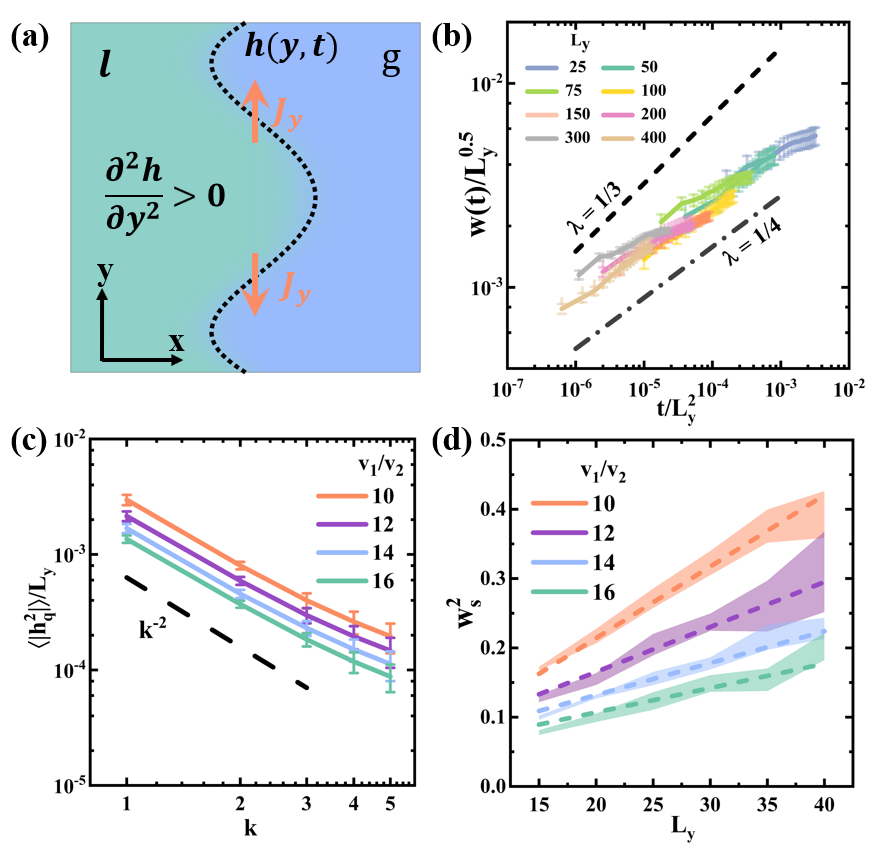}
\caption{(a) Schematic illustration of tangential fluxes caused by interfacial deformation, which restore the interfacial flatness through peak truncation and valley replenishment. (b) Temporal evolution of the interfacial width $w(t)$ from an initial flat interface ($v_1/v_2 = 10$ and $L_x/L_y = 2$). (c) Stationary fluctuations of the interface height for the flat interface with different $v_1/v_2$. (d) Steady-state squared interfacial width $\langle w_s^2\rangle$ as a function of $L_y$ ($L_x  = 100$ fixed), shown for different $v_1/v_2$. Here, the shaded areas are the simulation measurements with the corresponding standard deviation, and the dashed lines denote the slopes theoretically predicted from Eq.~(\ref{eq15}).}
\label{fig3}
\end{figure}

\emph{Dynamics of the MIPS interface}\textemdash To link the dynamics of interfacial height $h(y,t)$ to the density field evolution, we analyze the one-dimensional density profile along the $x$ direction at a fixed $y$ within the interfacial region. For a small interfacial slope $|\partial_y h| \ll 1$, it is  reasonable to assume the interfacial density profile remains invariant during the motion of the interface~\cite{bray2001interface,fausti2021capillary,langford2024theory,langford2025phase},
\begin{equation}\label{eq9}
\rho(\mathbf{r},t) = \psi(x - h(y,t)),
\end{equation}
where $\psi$ denotes the noise-averaged stationary interfacial density profile. Additionally, the conservation of particle number gives rise to,
\begin{equation}\label{eq10}
\int_\text{l}^\text{g} \frac{\partial \rho}{\partial t} dx = \frac{\partial h}{\partial t}(\rho_\text{l} - \rho_\text{g}) = - J_x^{\rho}|_\text{l}^\text{g} - \int_\text{l}^\text{g} \partial_y J_y^{\rho}dx.
\end{equation}
On the right-hand side of Eq.~(\ref{eq10}), the term of the $x$-direction flux (normal flux with respect to the average steady-state interface profile), which is governed solely by bulk phase properties and decoupled from the interfacial geometry, only provides a stochastic contribution to the interface dynamics. In contrast, as sketched in Fig.~\ref{fig3}(a), the tangential flux arises from the tangential ($y$-direction) density gradient induced by interface deformation and hence provides the restoring mechanism for the surface deformation.

To calculate the contribution of the tangential flux in Eq.~(\ref{eq10}), we utilize the condition of the invariant interfacial density profile (Eq.~(\ref{eq9})), which allows to express the tangential density gradient in terms of the normal density gradient, $\partial \rho/\partial y = -(\partial h/\partial y) (\partial \rho/\partial x)$. Using this relation and Eq.~(\ref{eq8}), Eq.~(\ref{eq10}) can be translated into the Langevin-type equation that governs the interfacial dynamics (see SM, Sec.~2.1~\cite{SupplRef} for a detailed derivation),
\begin{equation}\label{eq11}
\frac{\partial h(y,t)}{\partial t} = \alpha \frac{\partial^2 h(y,t)}{\partial y^2} + \zeta(y,t).
\end{equation}
Here, $\zeta(y,t)$ is a Gaussian white noise of zero mean and variance $\langle \zeta(y,t)\zeta(y',t') \rangle = \beta \delta(y - y') \delta (t - t')$. The coefficients $\alpha$ and $\beta$ in Eq.~(\ref{eq11}) are determined entirely by the bulk properties and the velocity function $v(\rho)$:
\begin{align}\label{eq12}
\alpha &= \frac{1}{4D_\text{r}(\rho_\text{g} - \rho_\text{l})} \left\{\left. \left[ v_0^2(\rho)\rho \right]\right|_\text{l}^\text{g} + \int_\text{l}^\text{g} v_0^2(\rho) d\rho \right\},\\
\beta &= \frac{1}{D_\text{r}(\rho_\text{g} - \rho_\text{l})^2}\left[ v_0^2(\rho_\text{g})\rho_\text{g} + v_0^2(\rho_\text{l})\rho_\text{l}\right]. \notag
\end{align}

Equation~(\ref{eq11}) indicates that the interfacial dynamics of the QS active system belongs to the Edwards-Wilkinson (EW) universality class~\cite{edwards1982surface}, instead of the Kardar-Parisi-Zhang (KPZ) one~\cite{kardar1986dynamic}. The EW model has previously been used to describe the interface dynamics of the MIPS of the conventional ABPs~\cite{lee2017interface,patch2018curvature,fausti2021capillary,langford2024theory}. As shown Sec.~2.4 of SM~\cite{SupplRef}, the EW model predicts that the transient interfacial width follows a power-law temporal scaling, $w(t) = \sqrt{\langle h^2(y,t) \rangle - \langle h(y,t)\rangle^2 } \propto t^\lambda$ with $\lambda = 1/4$, whereas for the KPZ model $\lambda = 1/3$. This prediction is well verified from the short-time simulations initialized with a flat interface, as shown in Fig.~\ref{fig3}(b).

The curvature term $\alpha\partial_y^2 h$ ($\alpha > 0$) in Eq.~(\ref{eq11}), which originates from the tangential flux, acts as a restoring force for the surface deformation. Its effect becomes more evident in the frequency domain. Given the interface's periodicity in the $y$-direction, we apply a Fourier transform to Eq.~(\ref{eq11}),
\begin{equation}\label{eq13}
\frac{\partial \tilde{h}(q_k,t)}{\partial t} = -\alpha q_k^2 \tilde{h}(q_k,t) + \tilde{\zeta}(q_k,t),
\end{equation}
with 
\begin{align}
\tilde{h}(q_k,t) &= \frac{1}{L_y}\int_0^{L_y} h(y,t)e^{-iq_ky} dy\notag\\
\tilde{\zeta}(q_k,t) &= \frac{1}{L_y}\int_0^{L_y} \zeta(y,t)e^{-iq_ky} dy,\notag
\end{align}
with the wavevector $q_k = 2\pi k/L_y$ for integer $k$. For $k \neq 0$, the stationary fluctuation of $\tilde{h}(q_k,t)$ reads (as detailed in the SM, Sec.~2.3~\cite{SupplRef}):,
\begin{equation}\label{eq14}
\lim_{t\rightarrow \infty}\langle |\tilde{h}(q_k,t)|^2 \rangle = \frac{\beta}{2\alpha L_y}\frac{1} {q_k^2}, %= \frac{\beta }{8\pi^2\alpha}\frac{L_y}{k^2}
\end{equation}
which dictates a $k^{-2}$ scaling behavior, consistent with the simulation results plotted in Fig.~\ref{fig3}(c).

Further, according to the capillary wave theory, we introduce the surface stiffness as $\kappa = (\lim_{t\rightarrow \infty}\langle |\tilde{h}(q_k,t)|^2 \rangle L_y q_k^2)^{-1}=2\alpha/\beta$, which describes the ability of the interface to resist deformation when subjected to perturbations. With the $\kappa$, the steady-state mean squared interface width $\langle w_s^2 \rangle = \lim_{t\rightarrow \infty}\langle w^2(t) \rangle$, which quantifies the magnitude of interface roughness, can be written as (see SM, Sec.~2.5~\cite{SupplRef} for details), 
\begin{equation}\label{eq15}
\langle w_s^2 \rangle = w_0^2 + \frac{L_y}{12\kappa}.
\end{equation}
Equation~(\ref{eq15}) provides us a simple way to directly compare the surface stiffness of the simulated QS system with the theoretical one predicted from Eq.~(\ref{eq12}). Figure~\ref{fig3}(d) shows that the measured $\langle w_s^2 \rangle$ from simulations quantitatively aligns with Eq.~(\ref{eq15}) with the theoretically predicted $\kappa$. This good agreement strongly supports the validity of the above theory.

Finally, it is interesting to differentiate the capillary surface tension $\gamma_\text{cw}$ of the QS system from the mechanical one $\gamma$. For interfaces in equilibrium, $\gamma_\text{cw}$ is related to the surface stiffness by $\gamma_\text{cw}=\kappa k_BT$ (with $T$ thermal bath temperature) and is identical to $\gamma$. However, in active system, the relationship between $\kappa$ and $\gamma_\text{cw}$ becomes amibiguous due to the presence of multiple distinct effective temperatures. Despite this fact, an effective capillary surface tension can still be defined for the current QS system as $\gamma_\text{cw}=\kappa k_BT_{\tt eff}$. Regardless of the selection of a reasonable effective temperature $T_{\tt eff}$, the effective $\gamma_\text{cw}$ remains nonzero, thereby essentially distinct from the vanishing mechanical surface tension. On the other hand, for a more general MIPS system with nonzero $\gamma$, it is natural to expect that $\gamma$ will also be responsible to keep the interface stability (as in passive systems), thus contributing to the effective $\gamma_\text{cw}$.

\emph{Conclusion}\textemdash We have investigated the unique interfacial behavior of ideal QS systems undergoing the MIPS. Our results reveal that in such MIPS systems, the mechanical surface tension vanishes completely\textemdash with interfacial mechanical equilibrium being sustained by polarization forces\textemdash while a substantial tangential-flux-induced surface stiffness (hence nonzero effective capillary tension) governs interfacial dynamics and stability. These findings underscore the fundamental difference between the mechanical and capillary surface tensions in active matter and significantly advance our understanding of active interfaces. This work sets the stage for further exploration of the interfacial behaviors in more realistic and intricate active systems.

\section*{Acknowledgment}
This work was supported by the National Natural Science Foundation of China (No. T2325027, 12274448, 12325405, 12090054, T2221001).

\bibliographystyle{apsrev4-2-titles}
\bibliography{Quorum_Sensing_Systems.bib}

\end{document}